\documentclass[%
 reprint,
superscriptaddress,
 amsmath,amssymb,
 aps,
floatfix,
]{revtex4-2}
\usepackage{footnote}
\usepackage{graphicx}
\usepackage{epstopdf}
\usepackage{bm}
\usepackage{mathtools}
\usepackage{textcomp}
\usepackage{amsmath}
\usepackage{physics}
\usepackage{float}
\usepackage{siunitx}
\usepackage{amssymb}
\usepackage{float}
\usepackage{steinmetz}
\usepackage{csquotes}
\usepackage{makecell}
\usepackage{accents}
\usepackage{lipsum}
\usepackage{bbold}
\setlipsum{%
	par-before = \begingroup\color{lightgray},
	par-after = \endgroup
}
\usepackage{soul,color}
\makeatletter
\def\maketitle{
	\@author@finish
	\title@column\titleblock@produce
	\suppressfloats[t]}
\makeatother

\usepackage{etoolbox}
\newcommand{\zerodisplayskips}{%
  \setlength{\abovedisplayskip}{0pt}%
  \setlength{\belowdisplayskip}{0pt}%
  \setlength{\abovedisplayshortskip}{0pt}%
  \setlength{\belowdisplayshortskip}{0pt}}
\appto{\normalsize}{\zerodisplayskips}
\appto{\small}{\zerodisplayskips}
\appto{\footnotesize}{\zerodisplayskips}

\newcommand\tab[1][0.8cm]{\hspace*{#1}}

\begin{document}
\preprint{APS/123-QED}

\title{Plasmonically engineered nitrogen-vacancy spin readout\vspace{-0.70em}}

\author{Harini Hapuarachchi}
\affiliation{ARC Centre of Excellence in Exciton Science and Chemical and Quantum Physics, School of Science, RMIT University, Melbourne, 3001, Australia}

\author{Francesco Campaioli}
\affiliation{Padua Quantum Technologies Research Center, Universita degli Studi di Padova, I-35131 Padua, Italy}

\author{Fedor Jelezko}
\affiliation{Institute for Quantum Optics, Ulm University, Albert-Einstein-Allee 11, 89081 Ulm, Germany}

\author{Jared H. Cole}
\affiliation{ARC Centre of Excellence in Exciton Science and Chemical and Quantum Physics, School of Science, RMIT University, Melbourne, 3001, Australia}

\date{\today}

\begin{abstract}\vspace{-1.0em}
Ultra-precise readout of single nitrogen-vacancy (NV) spins hold promise for major advancements in quantum sensing and computing technologies. We predict significant brightness and contrast enhancements in NV spin qubit readout and optically detected magnetic resonance (ODMR) arising from plasmonic interaction. We present a rigorous theory verified using existing measurements in the literature for such predictions. Plasmonic spin readout enhancements selectively manifest in carefully engineered parameter regions, necessitating rigorous modelling prior to experimentation.
\end{abstract}

\maketitle
Achieving high signal-to-noise ratio (SNR) and contrast (visibility) is crucial for realising spin-based sensors and measuring qubits at the quantum limit. In order to enhance optical readout in these applications, every photon counts. Therefore any techniques for increasing photon efficiency are invaluable.

\begin{figure}[t!]
	\includegraphics[width=0.9\columnwidth]{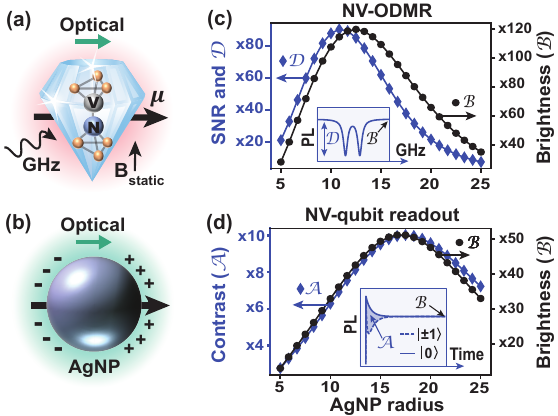} 
	\centering
	\caption{Conceptual illustrations of (a) an NV centre in a nanodiamond (ND) and (b) a plasmonic nanoparticle. (c) and (d) depict enhancements achievable in optically detected magnetic resonance (ODMR) and spin qubit readout by placing a single NV-bearing ND near a spherical silver nanoparticle in a high refractive index medium. All enhancements are relative to the isolated ND kept in the air/free space. The insets in subplots (c) and (d) depict the typical variations of ODMR and time-resolved spin qubit readout signals, respectively. The NV centre and plasmonic particle are assumed to be oriented such that their optical dipoles align, as depicted in Fig.~\ref{Fig:ODMR}(b). SNR enhancements in (c) are estimated by ODMR depth enhancements assuming that the associated noise power is constant. Parameters of (c) and (d) are the same as those of Fig.\ \ref{Fig:ODMR}(d) and Fig.\ \ref{Fig:TD_radial}, respectively.}\label{Fig:ToC_Fig}
\vspace{-0.4em}
\end{figure}

The negatively charged nitrogen-vacancy (NV) centre in diamond \cite{doherty2013nitrogen, rondin2014magnetometry, schirhagl2014nitrogen, aharonovich2011diamond}, conceptually depicted in Fig.\ \ref{Fig:ToC_Fig}(a), is a key building block in emerging room temperature quantum technologies. This is due to its robust optical and spin properties, high photostability, and biocompatibility. The detection of electron paramagnetic resonance (EPR) from single NV centres in 1997 by Gruber et al.\ \cite{gruber1997scanning} triggered intense research efforts towards their application in quantum information science \cite{wrachtrup2006processing, neumann2010quantum, jelezko2004observation, oberg2019spin, balasubramanian2009ultralong, childress2013diamond, lee2017topical, hahl2022magnetic, awschalom2018quantum, atature2018material}, sensing and metrology \cite{maletinsky2012robust, allert2022microfluidic, taylor2008high, dolde2011electric, doherty2014temperature, grinolds2013nanoscale, hall2009sensing, pelliccione2016scanned, doherty2014electronic, plakhotnik2014all, zhang2021diamond, nunn2022beauty, holzgrafe2020nanoscale, jeske2017stimulated, ajoy2015atomic, waddington2017nanodiamond} in diverse contexts, some of which are now undergoing commercialisation \cite{liu2022nitrogen, maletinsky2018n, aslam2023quantum}.

However, NV quantum technologies universally suffer from limitations of optical readout \cite{dolan2014complete}. For example, the fluorescence yield of NV centres can be as low as 5\% depending on the local environment \cite{plakhotnik2018nv}, and a multitude of factors impede their visibility \cite{liu2019coherent, doherty2013nitrogen, barry2020sensitivity, sengottuvel2022wide}. The diffraction limit \cite{stockman2008spasers} presents an additional barrier on enhancing the optical readout of quantum devices below microscale.

Recent experiments \cite{schietinger2009plasmon, bogdanov2018ultrabright} have demonstrated the possibility of enhancing the purely optical photoluminescence (PL) of NV centres by orders of magnitude beyond the diffraction limit, using plasmonic nanoparticles illustrated in Fig.\ \ref{Fig:ToC_Fig}(b).
Such enhancements stem from the plasmonically induced stronger optical transitions in the NV centre. While the intuition behind this is simple, the mechanism is complex and highly dependent on the relative arrangement, structure, material, background and inputs of the components. 
A verified derivation of NV-plasmonic optical interaction capturing these subtleties was introduced in our recent paper \cite{hapuarachchi2022nv}.

What is not directly intuitive is that plasmonic interaction can also enhance and control the contrast of NV spin readout in both frequency and time domains. This ability stems from the intricate interplay between the intrinsic nonradiative and the plasmonically modified radiative transitions of the NV centre. The necessity to simultaneously fine-tune the large NV-plasmonic parameter space has limited the experimental realisation of such enhancements to date.

Here we present a rigorous theory capable of capturing optical, vibronic, and spin interactions of the NV centre, both in the presence and absence of plasmonic interaction, verified against existing measurements in the literature. We predict significant readout strength and visibility enhancements for single NV spins in magnetometers and qubits, plasmonically achievable in carefully tuned parameter regions, as shown in Fig.~\ref{Fig:ToC_Fig}. 
Our theory equips the community with a tool to plasmonically engineer NV-spin readout enhancements in experimentation and device design.
\vspace{-2.2em}

\subsection*{Summary and verification of the formalism} \vspace{-1.5em}

\begin{figure}[ht!]
	\includegraphics[width=\columnwidth]{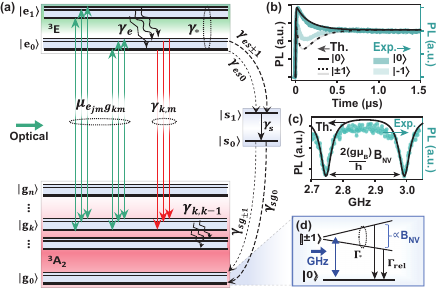}
	\centering
	\caption{NV model and verification. (a) The proposed NV$^-$ abstraction comprises $n+1$ ground optical/vibronic states $\lbrace \ket{g_k} \rbrace$ with energies $\lbrace\hbar\omega_k\rbrace$ ($k\in\lbrace 0, \hdots, n\rbrace$) and two optical excited states $\ket{e_0}$ and $\ket{e_1}$. The state $\ket{e_0}$ resides at the zero-phonon line (ZPL) energy $\hbar\omega_z$ with respect to the lowest ground state, and $\ket{e_1}$ is an effective upper excited level resonant with the photon energy of the incoming optical field $\hbar\omega_d$. The set of optical/vibronic states $\lbrace\ket{g_0}, \hdots, \ket{g_k}, \ket{e_0}, \ket{e_1}\rbrace$ are spin triplets of the form depicted in the lower right inset (d), interacting with microwave (GHz) and static magnetic fields along the NV axis ($B_\text{NV}$). $\ket{s_1}$ and $\ket{s_0}$ are spin singlet states where $\ket{s_0}$ is metastable. Other parameters shown include; dipole elements of optical transitions $\mu_{e_{jm}g_{km}}$, optical decay rates $\gamma_{k.m}$, ultrafast vibronic decay rate in the excited state $\gamma_e$, excited state dephasing rate $\gamma_*$, intersystem crossing rates to and from the singlets ($\gamma_{es\pm1}$, $\gamma_{es0}$, $\gamma_{sg\pm1}$, and $\gamma_{sg0}$), vibronic decay rates between adjacent ground states $\gamma_{k,k-1}$, as well as spin dephasing and spin relaxation rates ($\Gamma^*$, $\Gamma_\text{rel}$). (b) Time domain photoluminescence (PL) of an NV centre initialised in spin 0 and $\pm 1$ states obtained using the model, compared against the experimental data extracted from Steiner et al.\ \cite{steiner2010universal} (at the fitted optical field intensity $I\sim\SI{30}{\milli\watt\per\micro\meter\squared}$). (c) Comparison of our model with ODMR data extracted from Schirhagl et al.\ \cite{schirhagl2014nitrogen} (fitted microwave field amplitude $\sim\SI{0.35}{\milli\tesla}$ and $I \sim \SI{0.5}{\milli\watt\per\micro\meter\squared}$).}\label{Fig:Schematic_and_Verification}
\end{figure}

We first generalise the optical abstraction of the NV centre introduced in our previous work \cite{hapuarachchi2022nv}, accounting for the spin-related transitions, as depicted in Fig.~\ref{Fig:Schematic_and_Verification}(a). Each optical-vibronic level in $^3$E  and $^3$A$_2$ states expands into a 3-dimensional spin manifold $\lbrace\ket{+1}, \ket{0}, \ket{-1}\rbrace$, as depicted in Fig.~\ref{Fig:Schematic_and_Verification}(d). The lower and upper singlet states $\lbrace\ket{s_0}, \ket{s_1}\rbrace$ reside in a separate 2-dimensional subspace.

Our model abstractly mimics the operation of a single NV centre as follows:\ Optical radiation with energy higher than the ZPL drives spin-preserving coherent transitions between the vibronic levels of optical ground and excited states. Upon spin-preserving rapid vibronic relaxation of the excess energy in the excited state, decay of the orbital excitation could occur via two competing pathways. One is radiative decay into a ground vibronic state $\ket{g_k}$ while contributing a red photon to the $k^\text{th}$ band of the NV emission spectrum. This is followed by spin-preserving ground state vibronic transitions $\ket{g_k}\to\ket{g_{k-1}}$ until reaching the lowest state $\ket{g_0}$. The other decay pathway is to undergo intersystem crossing (ISC) via the singlet states $\ket{s_1}$ and $\ket{s_0}$. This pathway flips the spin state ($\ket{\pm1}\to\ket{0}$ and vice versa) of the NV centre. Orbital excitations in the $\ket{\pm1}$ states exhibit stronger ISC compared to $\ket{0}$. This leads to reduction of fluorescence when nearly resonant microwave ($\sim$GHz) fields coherently populate these spin-excited states. The spin $\ket{\pm1}$ states further undergo Zeeman splitting proportional to static magnetic fields ($B_\text{\tiny{NV}}$) incident along the NV axis, enabling measuring $B_\text{\tiny{NV}}$ by optically detecting the microwave spin resonance.

Accounting for the above interactions, and utilising the rotating wave approximation, we derive the following Hamiltonian for the NV centre interacting with optical, microwave, and static magnetic fields: 
\begin{align}
&H_\text{\tiny{NV}}
\approx 
\bigg\lbrace\small{\sum}_k\ket{g_k}\bra{g_k} \otimes H_\text{gs} + \small{\sum}_j\ket{e_j}\bra{e_j} \otimes H_\text{es}\\[-0.3em]
&+\small{\sum}_k \left\lbrace \hbar\omega_k \ket{g_k}\bra{g_k}\otimes\mathbb{1}_3 \right\rbrace + \hbar(\omega_z-\omega_\text{d})\ket{e_0}\bra{e_0}\otimes\mathbb{1}_3 \nonumber\\[-0.3em]
&-\small{\sum}_{jkm}\hbar\Omega_{e_{jm}g_{km}} (\ket{e_j}\otimes\ket{m})(\bra{g_k}\otimes\bra{m}) + h.c.\bigg\rbrace \oplus H_s\nonumber
\end{align}
where the high-frequency time dependence in the orbital and spin subsystems have been eliminated via unitary transformations. For an optical field $E = E_0 (e^{-i\omega_\text{d} t} + c.c.)$ incident along the plane of optical dipole formation of the NV centre \cite{hapuarachchi2022nv}, the Rabi frequency of the coherent transition  $\ket{e_j}\otimes\ket{m}\leftrightarrow\ket{g_k}\otimes\ket{m}$ is denoted by $\Omega_{e_{jm}g_{km}} = \Omega^0_{e_{jm}g_{km}} = \mu_{e_{jm}g_{km}}E_0/\left(\hbar\epsilon_\text{effD}\right)$, where $\omega_\text{d}$ and $E_0$ are the angular frequency and positive frequency amplitude of the optical field, $\mu_{e_{jm}g_{km}}$ is the optical dipole element of the NV transition. The factor  $\epsilon_\text{effD}$ accounts for the optical field screening induced by the emitter material (diamond). $H_\text{gs}$, $H_\text{es}$, and $H_s$ denote Hamiltonians of spin manifolds in the orbital ground and excited states (which capture the microwave and Zeeman interactions) and the singlet subsystem, respectively. 

When a plasmonic nanoparticle much smaller than the wavelength of incoming light is placed at a nanoscale separation, it undergoes dipolar interactions with the NV centre and the incoming optical field. It modifies the Rabi frequency of the optical transitions $\ket{e_j}\otimes\ket{m}\leftrightarrow\ket{g_k}\otimes\ket{m}$ and optical emission rates $\gamma_{k,m}$ (in Fig.~\ref{Fig:Schematic_and_Verification}(a)), while absorbing a fraction of the photons emitted by the NV centre. We assume that adjacent plasmonic excitations do not adversely impact the microwave and magnetic field interactions of the NV centre, judging by the recent experimental demonstrations of plasmonically enhanced ODMR with spin defects in hBN \cite{gao2021high, xu2022greatly, mendelson2022coupling}. We further assume that magnetic fields involved in NV-based ODMR are too small to induce any significant Faraday rotation of the optically induced plasmonic dipole \cite{mazor2017dark}.

\begin{figure*}[t!]
	\includegraphics[width=0.90\textwidth]{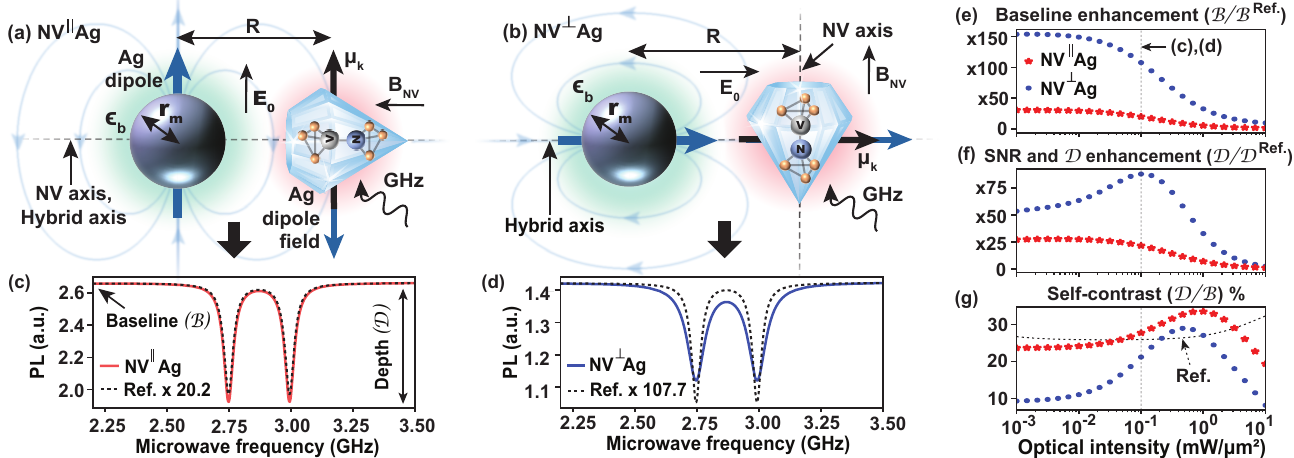}
	\centering
	\caption{Enhancing the visibility of ODMR signals from single NV centres using plasmonic nanoparticles. (a) Schematic illustration of the NV$^\parallel$Ag setup where the external optical field ($E_0$) polarisation is perpendicular to the hybrid axis and NV dipoles $\mu_k$ (assuming $\mu_{e_{jm}g_{km}} = \mu_k \;\forall j,m$) are parallel to the surface of the Ag nanoparticle. (b) NV$^\perp$Ag setup where the external optical field polarisation is along to the hybrid axis and NV dipoles $\mu_k$ form perpendicular to the surface of the Ag nanoparticle. In (a) and (b), a static magnetic field $B_\text{\tiny{NV}}\approx\SI{4.4}{\milli\tesla}$ is incident along the NV axis and the magnetic component of the microwave (GHz) field ($\approx\SI{0.35}{\milli\tesla}$) is assumed to be along an arbitrary direction perpendicular to the respective NV axis. The red and blue solid lines in (c) and (d) show ODMR spectra of NV centres in NV$^\parallel$Ag and NV$^\perp$Ag configurations, respectively, for an optical excitation intensity $I=\SI{0.1}{\milli\watt\per\micro\meter\squared}$. The dashed black lines in (c) and (d) depict the ODMR spectrum of the reference NV centre at the same intensity, scaled by the baseline enhancement of the respective NV$^\measuredangle$Ag setup ($\measuredangle\in\lbrace\parallel, \perp\rbrace$). (e) and (f) depict the variation of ODMR baseline and depth enhancements of NV centres in NV$^\parallel$Ag and NV$^\perp$Ag setups, in comparison to the reference, plotted against the intensity of the incident optical field. SNR enhancements can be estimated by depth enhancements when the associated noise power is assumed constant \cite{hopper2018spin}. (g) shows the ODMR contrast of the reference, NV$^\parallel$Ag, and NV$^\perp$Ag setups as a function of the incoming optical field intensity. The legend in (e) is common for (e-g). For all sub-figures, $r_\text{m} = \SI{10}{\nano\meter}$, $R=r_\text{m}+\SI{10}{\nano\meter}$, $\epsilon_\text{b}=n_\text{b}^2$ where $n_\text{b}\sim 2.426$, $\hbar\omega_\text{d}\sim \SI{2.033}{\electronvolt}$ (equivalent to a free-space wavelength $\approx \SI{532}{\nano\meter}$),  and reference (Ref) is a single NV centre in an isolated nanodiamond in free-space.}\label{Fig:ODMR}
\end{figure*}

We model the evolution of NV density operator $\rho$ using the Lindblad formalism \cite{campaioli2023tutorial}, incorporating the coherent interactions captured in the Hamiltonian and all decoherence mechanisms depicted in Fig.~\ref{Fig:Schematic_and_Verification}, both in the presence and absence of plasmonic excitations. 
We then simulate the ODMR spectra by estimating the photoluminescence (PL) counts at a given microwave frequency as follows,
\begin{equation}\label{Eq:PL}
	\text{PL} \approx \small{\sum}_{k} \small{\sum}_{m} Q_{k,m}\gamma_{k,m}\rho_{e_{0m}e_{0m}}.
\end{equation}
$\rho_{e_{0m}e_{0m}}$ is the diagonal element of $\rho$ denoting the probability of finding the NV centre in $m^\text{th}$ spin state of $\ket{e_0}$. $Q_{k,m}$ is the fraction of $\ket{e_0}\otimes\ket{m}\to\ket{g_k}\otimes\ket{m}$ emissions surviving in the far field following plasmonic absorption. $Q_{k,m} = 1$ in the absence of plasmonic interactions.

We have verified this formalism against existing measurements in the literature as depicted in Fig.\ \ref{Fig:Schematic_and_Verification}.
Further verifications, both in the presence and absence of plasmonic nanoparticles, and details of mathematical abstractions are provided in the supplementary material.
\vspace{-2.0em}

\subsection*{NV plasmonics} \vspace{-1.2em}
In this work, we consider optically illuminating an NV-Ag nanoparticle (AgNP) pair near the plasmonic resonance of the AgNP, which significantly enhances the Rabi frequencies of the NV optical transitions. 

Plasmonic resonance of a spherical AgNP in air typically resides in the (far violet or) ultraviolet (UV) frequency range ($\sim$$\SI{3.5}{\electronvolt}$) \cite{hapuarachchi2022nv, maier2007plasmonics}. It has been shown that such high-frequency illumination is likely to result in PL intensities much weaker compared to those resulting from green illumination. This occurs predominantly via the excitation of NV$^0$ charge state instead of the expected NV$^-$ in such regimes \cite{yang2022photoluminescence, lu2019nitrogen}. To achieve the desired response, we assume a high-refractive index embedding medium capable of red-shifting \cite{maier2007plasmonics, hapuarachchi2017cavity} the Ag plasmonic resonance towards the green region. Such embedding media could include diamond itself. 
Embedding AgNPs in diamond has previously been reported in the literature \cite{li2018enhanced, li2019silver, ramsay2022nanostructured}. These include a demonstration of enhanced emission from an SiV centre incorporated into the diamond host surrounding an AgNP \cite{li2019silver}. In addition to red-shifting the plasmonic peak, such embedding protects the AgNP against oxidation \cite{li2019silver,ramsay2022nanostructured}. Furthermore, high refractive index media cause amplification of plasmonic peak \cite{maier2007plasmonics, hapuarachchi2017cavity}, enabling stronger local field enhancements (hence faster NV Rabi frequencies) compared to those achievable in free-space. 

Alternative approaches to red-shifting the resonance include modifying the shape of the AgNP \cite{amendola2010study, ameer2016tuning}. 
For simplicity, in this letter, we consider spherical AgNPs. 

In Fig.\ \ref{Fig:ODMR}, we consider two extreme orientations of an NV-Ag pair relative to the applied optical field. PL modifications herein arise from the interplay between several factors including; Ag-induced optical Rabi frequency enhancements and decay rate modifications \cite{hapuarachchi2022nv}, partial absorption of NV emission by the AgNP \cite{carminati2006radiative}, and the reduction of optical screening experienced by the NV centre \cite{artuso2012optical}. Despite favouring PL amplification, plasmon-induced decay rate enhancement tends to compromise the relative contrast of ODMR signals.
This is due to the enhanced radiative emission diminishing the excited state $m=\pm 1$ population available to nonradiatively decay via the intersystem crossing (ISC) pathway. Here we assume that the intrinsic ISC rates of the NV centre remain unaffected by the AgNP.

When submerged in diamond-like media, the dipolar plasmonic response of the AgNP follows a steep Lorentzian-like line shape peaking around green optical energies. Therefore, AgNP-induced emission rate enhancement in the far detuned red emission region of the NV centre is lower compared to enhancements attainable near the plasmon resonance. This, accompanied by the strong Rabi enhancements achieved by optically illuminating at the plasmonic peak, result in strong baseline enhancement without compromising the visibility of the resulting ODMR signals. Significant detuning of the NV emission from the plasmonic peak also allows retaining a reasonable proportion of PL in the far field for detection, following plasmonic absorption. Rabi and decay rate enhancement spectra and relative quantum efficiency estimates for our example NV-Ag setups can be found in the supplementary material. The reference (ref.) mentioned throughout this work is an NV centre embedded in a nanodiamond kept in the free-space.
\vspace{-2.0em}

\subsection*{Plasmonically enhanced NV-ODMR} \vspace{-1.5em}

In Fig.\ \ref{Fig:ODMR}, we introduce three figures of merit (FoMs), namely, baseline enhancement, depth enhancement, and self-contrast to quantify the efficacy of plasmonic interaction in enhancing dc magnetometry of single NV centres. Our results suggest that these FoMs can be preferentially enhanced in carefully tuned parameter regions.

In the example NV$^\parallel$Ag setup considered in Fig.\ \ref{Fig:ODMR}(c), the NV centre experiences $\sim$20 fold ODMR baseline enhancement compared to the reference. In the respective NV$^\perp$Ag configuration in Fig.\ \ref{Fig:ODMR}(d), more than $\sim$100 fold ODMR baseline enhancement is observed, at an apparent compromise of the relative ODMR contrast compared to the reference. However, it is noteworthy from Fig.\ \ref{Fig:ODMR}(f), that the absolute ODMR depth is enhanced $\sim$80 fold in comparison to the reference in this case. 

We estimate the photon shot-noise-limited dc magnetic field sensitivity in ODMR as \cite{dreau2011avoiding, barry2016optical}, $\eta_\text{B} = \delta B_\text{min}\sqrt{\Delta t}\sim 4 h\Delta\nu/(3\sqrt{3}g \mu_B C \sqrt{\text{PL}_\text{m}})$ where $\Delta\nu$, $C$, and $\text{PL}_\text{m}$ denote the resonance linewidth, fractional contrast, and photon detection rate away from resonance (obtained using (\ref{Eq:PL})). The minimum detectable magnetic field within a measurement duration $\Delta t$ is denoted by $\delta B_\text{min}$ \cite{dreau2011avoiding}. The NV$^\parallel$Ag and NV$^\perp$Ag setups in Fig.\ \ref{Fig:ODMR}(c) and (d) possess $\eta\sim$$\SI{2.7}{\micro\tesla\per\sqrt{\hertz}}$ and $\eta\sim$$\SI{2.5}{\micro\tesla\per\sqrt{\hertz}}$ respectively, both of which amount to $\sim$5-fold improvements compared to the reference sensitivity of $\sim$$\SI{12.2}{\micro\tesla\per\sqrt{\hertz}}$ (computed for an isolated NV-bearing nanodiamond kept in free-space).


\begin{figure}[ht!]
	\includegraphics[width=\columnwidth]{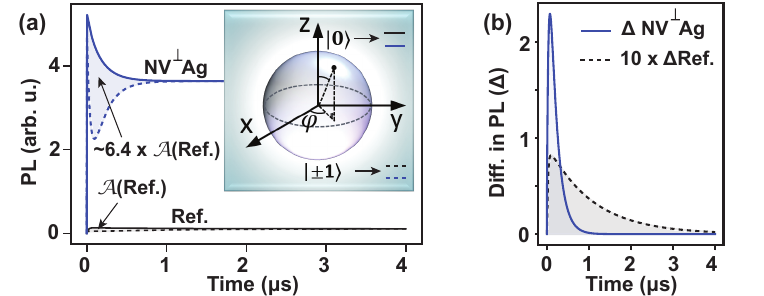}
	\centering
	\caption{Time domain spin contrast comparison. (a) Time domain PL counts from the NV centre in the NV$^\perp$Ag setup in Fig.\ \ref{Fig:ODMR}(b) for spin initialisations in $\ket{0}$ (solid blue line) and $\ket{\pm1}$ (dashed blue line) states. The black solid and dashed lines depict the same for an isolated nanodiamond in air hosting a single NV centre (Ref). (b) The difference $(\Delta)$ between respective signals for $\ket{0}$ and $\ket{\pm1}$ spin initialisations in the previous subplot. The $\Delta$ signal for the reference ($\Delta\text{Ref.}$) is magnified 10 times for better visibility. $I\sim\SI{1}{\milli\watt\per\micro\meter\squared}$. 
    } \label{Fig:TD_radial}
\end{figure}

It is evident from Fig.\ \ref{Fig:ODMR}(e), that ODMR baseline enhancements of both NV$^\parallel$Ag and NV$^\perp$Ag setups decline with increasing optical intensities, which is likely to be attributable to the saturation of the NV centre. For weak optical intensities of the scale $\SI{1}{\micro\watt\per\micro\meter\squared}$, baseline enhancements reach $\sim$30- and $\sim$150- fold for the two extreme setups, NV$^\parallel$Ag and NV$^\perp$Ag, respectively. It is observable from Fig. \ref{Fig:ODMR}(f), that these are accompanied by $\sim$25- and $\sim$50- fold enhancements of the respective absolute ODMR depths, in comparison to the reference. Baseline and depth enhancements expectable for intermediate orientations are likely to reside within these bounds. Therefore these results show prospects of using NV-Ag setups embedded in high refractive index media (such as Ag-diamond core-shell particles implanted with NV centres) for sensing applications deeper within environments with relatively low optical penetration. 
\vspace{-1.5em}

\subsection*{Plasmonically enhanced NV spin qubit readout} \vspace{-1.2em}

In Fig.\ \ref{Fig:TD_radial}, we predict plasmonically enhancing the time domain optical spin readout of NV centres, using the NV$^\perp$Ag configuration as an example. When initialised in $m=0$ state, the larger Rabi frequency and radiative decay rate enhancements experienced by the NV centre in NV$^\perp$Ag setup result in an initially high level of PL. This level declines as the metastable state populates and eventually stabilizes as observable in Fig.\ \ref{Fig:TD_radial}(a). When the same NV centre is initialised in the $m=\pm 1$ states which exhibit stronger ISC rates compared to the brighter $m=0$ state, a larger population quickly reaches the metastable state, declining the early PL levels. The signal eventually recovers and reaches the steady state, as the metastable state excitations gradually populate the brighter $m=0$ spin state. 

{For the NV$^\perp$Ag setup in Fig.\ \ref{Fig:TD_radial}, we observe a $\sim$33-fold steady-state PL enhancement and a 6.4-fold spin contrast (area) enhancement, compared to the reference. The Ag-assisted signal stabilises $\sim$4 times faster than the reference, reducing the optical pulse length needed to determine the initial spin state of the NV centre, as observable in Fig.\ \ref{Fig:TD_radial}(b). Such improvements could lead to enhanced optical readout of NV qubit states in room-temperature quantum computing and communication applications. \vspace{-1.5em}

\subsection*{Summary and outlook} \vspace{-1.5em}

We present a verified open quantum system model that accounts for optical, vibronic, and spin transitions of the negatively charged nitrogen-vacancy (NV) centre, both in the absence and presence of plasmonic interactions. Utilising this model, we predict the possibility of achieving orders of magnitude enhancements of both baselines and dips of NV-based optically detected magnetic resonance (ODMR) signals using silver nanoparticles embedded in high permittivity media. Orders of magnitude enhancements of both time-domain readout contrast and brightness, accompanied by several-fold reduction of readout time, is predicted for NV-based spin qubits. These enhancements occur in precisely tuned parameter regions, necessitating theoretical modelling prior to experiment and device design aimed at harnessing them.

Diamond is tolerant of extremely harsh environmental conditions, biocompatible, and functionalisable with a range of different proteins and targeting ligands \cite{nunn2022beauty}. Therefore, such ODMR enhancements in NV-Ag hybrids encapsulated in diamond hold promise for advancing a myriad of timely applications. These include 
sub-cellular magneto-optical detection of substances and activity deeper within tissue \cite{nunn2022beauty}, mineral exploration \cite{ruan2018magnetically}, and navigation in extreme environments \cite{wang2023quantum}. The time domain enhancements hold significant promise for high-fidelity single-shot readout of spin qubits which is crucial for achieving fault-tolerant quantum computing and scalable quantum networks at room-temperature \cite{zhang2021high, ghassemizadeh2022stability}.

\subsection*{Acknowledgements} 
HH acknowledges Massimiliano Ramsay (London Centre for Nanotechnology), Miles I Collins (UNSW Sydney), Philipp J Vetter and Christian Osterkamp (Ulm University), and Dinuka U Kudavithana for insightful discussions. This work was funded by the Australian Research Council (grant number CE170100026) and the National
Computational Infrastructure (NCI), supported by the Australian Government.

\clearpage
%
%

\widetext
\begin{center}
	\textbf{\Large Supplemental Materials: \\Plasmonically engineered nitrogen-vacancy spin readout\\}\vspace{0.5em}
	{Harini Hapuarachchi$^1$, Francesco Campaioli$^2$, Fedor Jelezko$^3$, and Jared H. Cole$^1$}\\ \vspace{0.5em}
	$^1$\emph{ARC Centre of Excellence in Exciton Science and Chemical and Quantum Physics, \\School of Science, RMIT University, Melbourne, 3001, Australia}\\
	$^2$\emph{Padua Quantum Technologies Research Center, Universita degli Studi di Padova, I-35131 Padua, Italy}\\
	$^3$\emph{Institute for Quantum Optics, Ulm University, Albert-Einstein-Allee 11, 89081 Ulm, Germany}\\ \vspace{1.5em}
\end{center}
\twocolumngrid
%
%
%
%
%
%
\renewcommand{\theequation}{\arabic{equation}}
\setcounter{equation}{0}  



\section{Supporting information on the analytical formalism} \vspace{-1em}

\subsection{Hamiltonians of the spin triplet and singlet subsystems}\vspace{-1em}
In this work, we consider a case where coherent transitions can be driven in the NV ground and excited state spin manifolds by an oscillatory microwave magnetic field of the form, $$\undertilde{B} = \undertilde{B}_{\mu0}\left(e^{-i\omega_\mu t} + e^{i\omega_\mu t}\right),$$ with angular frequency $\omega_\mu$ and amplitude $B_{\mu0}$ aligned transverse to the NV axis. A static magnetic field $B_\text{\tiny{NV}}$ (a few mT or smaller) is incident along the NV axis. In the absence of hyperfine interactions, we estimate the Hamiltonians of NV ground (gs) and excited (es) state spin manifolds in a reference frame rotating at $\omega_\mu$ as,\vspace{1em}
\begin{subequations}
	\begin{align*}\vspace{1.5em}
		H_\text{gs} &= \hbar(\omega_\text{\tiny{g+1}} - \omega_\mu)\ket{+1}\bra{+1} +  \hbar(\omega_\text{\tiny{g-1}} - \omega_\mu)\ket{-1}\bra{-1} \\
		& \;\;\;\;\;\; + \hbar\Omega_\mu(\ket{0}\bra{+1} + \ket{0}\bra{-1} + h.c.) \\
		H_\text{es} &= \hbar(\omega_\text{\tiny{e+1}} - \omega_\mu)\ket{+1}\bra{+1} +  \hbar(\omega_\text{\tiny{e-1}} - \omega_\mu)\ket{-1}\bra{-1} \\
		& \;\;\;\;\;\; + \hbar\Omega_\mu(\ket{0}\bra{+1} + \ket{0}\bra{-1} + h.c.), 
	\end{align*}
\end{subequations}
where $\Omega_\mu = (g\mu_\text{\tiny{B}} B_{\mu 0})/\big(\hbar\sqrt{2})$ is the Rabi frequency of the coherent spin transitions, $g\approx 2.0$ is the Land\`{e} $g$ factor, $\mu_\text{\tiny{B}}$ is the Bhor magneton, and $\hbar$ is the reduced Planck's constant. The rotating wave approximation has been used to eliminate the high-frequency time dependence when deriving the above Hamiltonians. 

For the simplicity of analytics we assume that; any transverse static magnetic fields present are much smaller than $B_\text{\tiny{NV}}$, and any off-axial zero-field splitting parameters resulting from the local strain in the diamond lattice are much smaller than the axial zero-field splitting parameters $D_\text{gs}\approx\SI{2.87}{\giga\hertz}$ and $D_\text{es}\approx\SI{1.42}{\giga\hertz}$. With these assumptions, the angular eigenfrequencies of the $\ket{\pm1}$ spin sublevels relative to $\ket{0}$ in the above Hamiltonians can be estimated as \cite{rondin2014magnetometry},\vspace{0.75em}
\begin{subequations}
	\begin{align*}
	\omega_{\text{g}\pm1} = 2\pi D_\text{gs} \pm (g\mu_\text{\tiny{B}}/\hbar)B_\text{\tiny{NV}},\\
	\omega_{\text{e}\pm1} = 2\pi D_\text{es} \pm (g\mu_\text{\tiny{B}}/\hbar)B_\text{\tiny{NV}}.
	\end{align*}
\end{subequations}

Hamiltonian of the two-dimensional spin singlet subsystem $H_s$ is assumed to take the following form, in a frame rotating at $\omega_\text{d}$,
\begin{equation*}
	H_s = \hbar(\omega_{s_1} - \omega_\text{d})\ket{s_1}\bra{s_1} + \hbar(\omega_{s_0} - \omega_\text{d})\ket{s_0}\bra{s_0},
\end{equation*} 
where $\omega_{s_1}$ and $\omega_{s_0}$ denote the angular eigenfrequencies of the singlet states $\ket{s_1}$ and $\ket{s_0}$.
\vspace{-1em}

\subsection{Plasmonic modification of optical Rabi frequency}\vspace{-1em}
Extending the procedure established in our earlier work \cite{hapuarachchi2022nv}, the effective Rabi frequency experienced by the $\ket{e_j}\otimes\ket{m}\leftrightarrow \ket{g_k}\otimes\ket{m}$ coherent transitions of the NV centre in the presence of a proximal plasmonic nanoparticle can be estimated as,
\begin{equation*}\label{Eq:Rabi_freq}
	\Omega_{e_{jm}g_{km}} = \Omega^\text{pl}_{e_{jm}g_{km}} = \Omega^\text{lin}_{e_{jm}g_{km}} + \Omega^\text{nl}_{e_{jm}g_{km}}
\end{equation*}
where the linear and nonlinear (self-feedback induced) optical Rabi components take the following forms,
\begin{subequations}
	\begin{align*}
		\Omega^\text{lin}_{e_{jm}g_{km}} &= \frac{\mu_{e_{jm}g_{km}} E_0}{ \hbar \epsilon_\text{effD}}\left( 1 + \frac{s_\alpha\alpha(\omega_d)}{R^3} \right),\\
		\Omega^\text{nl}_{e_{jm}g_{km}} &= \eta_{e_{jm}g_{km}} \sum_{j'=0}^1\sum_{k'=0}^n\sum_{m'=-1}^{+1}\left(\mu_{e_{j'k'}g_{k'm'}} \tilde{\rho}_{e_{j'k'}g_{k'm'}}\right)\\
		\text{with,}&\;\; \eta_{e_{jm}g_{km}} =\frac{\mu_{e_{jm}g_{km}} s_\alpha^2 \alpha(\omega_d)}{\hbar(4\pi\epsilon_0\epsilon_\text{b})\epsilon_\text{effD}^2R^6}.
	\end{align*} 
\end{subequations}
For the simplicity of analyses, here we capture only two special cases where the external optical field $E = E_0 e^{-i\omega_d t} + c.c.$ is aligned such that the NV dipoles form either radially ($s_\alpha=2$) or tangentially ($s_\alpha=-1$) to the surface of an adjacent spherical plasmonic nanoparticle. In the above equations, $\epsilon_\text{b}$ and $\epsilon_\text{\tiny{D}}$ denote the relative permittivities of the background medium and diamond and $R$ denotes the centre separation between the NV centre and the plasmonic nanoparticle. The factor $\epsilon_\text{effD} = (2\epsilon_\text{b} + \epsilon_\text{\tiny{D}})\big/(3\epsilon_\text{b})$ captures the optical screening induced by the emitter material (diamond) \cite{hapuarachchi2022nv,artuso2012optical}. It arises from the solution of Laplace's equation for a homogenous isotropic particle with a linear dielectric placed in an electric field \cite{artuso2012optical}.

The dipolar polarizability of the spherical plasmonic nanoparticle at the angular frequency of the incoming optical field $\omega_\text{d}$ has been modelled as follows, while taking the finite-size effects into account \cite{colas2012mie, des2016plasmonic, carminati2006radiative, hapuarachchi2022nv}:
\begin{equation*}\label{Eq:alpha_eff}
	\alpha(\omega_\text{d}) = \frac{\alpha_\text{\tiny{L}}(\omega_\text{d})}{\left[1-\frac{2i k_\text{b}^3}{3}\alpha_\text{\tiny{L}}(\omega_\text{d})\right]}.
\end{equation*}
Here, $\alpha_\text{\tiny{L}}(\omega_\text{d}) = r_\text{m}^3\left[\epsilon_\text{m}(\omega_\text{d}) - \epsilon_\text{b}\right]\big/\left[\epsilon_\text{m}(\omega_\text{d}) + 2 \epsilon_\text{b}\right]$, $r_\text{m}$ denotes the plasmonic nanoparticle radius, and $\epsilon_\text{m}(\omega_\text{d})$ is its relative permittivity at $\omega_\text{d}$. Wavenumber of the non-absorbing, non-magnetic background medium is $k_\text{b} = n_\text{b}k$, where $n_\text{b} = \sqrt{\epsilon_\text{b}}$ is its refractive index, and $k=\omega_\text{d}/c$ is the free-space wavenumber ($c$ is the speed of light). 

Based on sampled analyses, it could be observed that $\Omega^\text{nl}$ is several orders of magnitude smaller than $\Omega^\text{lin}$ in the parameter region considered for this study. Therefore, the optical Rabi frequency was approximated as linear when generating the results presented.
\\

\noindent
\textbf{Regarding plasmon emission:} This model does not account for the radiative emission of plasmonic nanoparticles which could contribute photons to the detected counts \cite{schietinger2009plasmon}. Suggested approaches to eliminate any contributions from such emission include: filtering (as plasmonic peak is far off-resonant from the NV emission regime), or measuring the plasmonic emission counts prior to incorporating the NV centre to correct the NV emission spectra accordingly.
\vspace{-1em}

\subsection{NV centre as an open quantum system}\vspace{-1em}
Hamiltonian of the NV centre coupled to the external and plasmon-induced optical, microwave, and static magnetic fields represents a closed quantum system where the interactions with the environment are yet to be taken into account. We assume that the NV centre weakly couples to its environment resulting in an open quantum system with irreversible (incoherent) dynamics. We estimate the evolution of its density operator under the influence of a Markovian (memoryless) bath using a Lindblad master equation-based approach, as follows \cite{breuer2002theory, manzano2020short, campaioli2023tutorial},
\begin{multline*}\label{Eq:Density_matrix_master_eq}
	\dot{\rho} = -\frac{i}{\hbar}[H_\text{\tiny{NV}}, \rho] + \sum_{x}\Gamma_x[L_x^{\phantom{\dagger}}\rho L_x^\dagger -\frac{1}{2}\lbrace L_x^\dagger L_x^{\phantom{\dagger}}, \rho\rbrace].
\end{multline*}
In the above,  $\rho$ denotes the density operator of the NV centre in the same reference frame as the Hamiltonian $H_\text{\tiny{NV}}$. The mathematical operators $[\cdot,\cdot]$ and $\lbrace\cdot,\cdot\rbrace$ represent the commutator and anti-commutator of the operands. $L_x$ denotes the Lindblad or collapse operator corresponding to the $x$th decoherence channel with characteristic the decoherence rate $\Gamma_x$. 
\vspace{-1em}

\subsection{The list of collapse operators}\vspace{-1em}
The decoherence channels and the corresponding rates considered in our open quantum system model, schematically depicted in Fig.\ 1 of the main text, are listed below. In the following expressions, $\mathbb{0}_{n+3}$, $\mathbb{0}_{3}$, and $\mathbb{0}_{2}$ denote the zero operators and $\undertilde{0}_{n+3}$, $\undertilde{0}_3$, and $\undertilde{0}_2$ denote the all-zero (column) vectors of the optical-vibronic, spin, and singlet Hilbert subspaces, respectively.

\begin{flushleft}
	For each spin preserving optical emission transition:\\
	$\Gamma_x = \gamma_{k,m}$, for $k\in\lbrace 0,\hdots,n\rbrace$ and $m\in\lbrace +1,0,-1 \rbrace$\\
	$L_x = \sigma_{k,m} = \left(\ket{g_k}\otimes\ket{m}\right)\left(\bra{e_0}\otimes\bra{m}\right)\oplus\mathbb{0}_2$
\end{flushleft}

\begin{flushleft}
	For each spin-preserving vibronic decay transition in the $^3A_2$ ground state:\\
	$\Gamma_x = \gamma_{k,k-1}$, for $k\in\lbrace 1,\hdots,n\rbrace$ and $\forall m$\\
	$L_x = \left(\ket{g_{k-1}}\otimes\ket{m}\right)\left(\bra{g_k}\otimes\bra{m}\right)\oplus\mathbb{0}_2$
\end{flushleft}

\begin{flushleft}
	For spin-preserving vibronic decay in $^3E$ excited state:\\
	$\Gamma_x = \gamma_e$, $\forall m$\\
	$L_x = \left(\ket{e_0}\otimes\ket{m}\right)\left(\bra{e_1}\otimes\bra{m}\right)\oplus\mathbb{0}_2$
\end{flushleft}

\begin{flushleft}
	Dephasing due to the stochastic energy fluctuations of the $^3E$ excited state relative to $^3A_2$ ground state:\\
	$\Gamma_x = \gamma_*$\\
	$L_x = \left\lbrace \sum_{j=0}^1\sum_{m=-1}^{+1} (\ket{e_j}\otimes\ket{m})(\bra{e_j}\otimes\bra{m})\right\rbrace\oplus\mathbb{0}_2$
\end{flushleft}

\noindent
The spin preserving vibronic relaxation rates in the excited and ground levels  $\gamma_e$ and $\gamma_{k,k-1}$ reside in THz ranges, far exceeding all other rates of decoherence in the system. Therefore, we assume that it suffices to account for the spin decay and dephasing only in the triplet manifolds of the lowest energy excited and ground states $\ket{e_0}$ and $\ket{g_0}$ as follows:

\begin{flushleft}
	Spin relaxation in the excited state $^3E$:\\
	$\Gamma_x = \Gamma_\text{rel}^e$\\
	$L_x = \left(\ket{e_0}\otimes\ket{0}\right)\left(\bra{e_0}\otimes\bra{\pm1}\right)\oplus\mathbb{0}_2$
\end{flushleft}

\begin{flushleft}
	Spin relaxation in the ground state $^3A_2$:\\
	$\Gamma_x = \Gamma_\text{rel}^g$\\
	$L_x = \left(\ket{g_0}\otimes\ket{0}\right)\left(\bra{g_0}\otimes\bra{\pm1}\right)\oplus\mathbb{0}_2$
\end{flushleft}

\noindent
We account for spin dephasing arising due to stochastic energy fluctuations of $\ket{\pm1}$ states relative to $\ket{0}$ as follows (assuming opposite signs for the respective energies):
\begin{flushleft}
	Spin dephasing in the excited state $^3E$:\\
	$\Gamma_x = \Gamma_*^e$\\
	$L_x = \bigg\lbrace\left(\ket{e_0}\otimes\ket{+1}\right)\left(\bra{e_0}\otimes\bra{+1}\right) -$\\ \tab\tab\tab\tab  $\left(\ket{e_0}\otimes\ket{-1}\right)\left(\bra{e_0}\otimes\bra{-1}\right) \bigg\rbrace\oplus\mathbb{0}_2$
\end{flushleft}

\begin{flushleft}
	Spin dephasing in the ground state $^3A_2$:\\
	$\Gamma_x = \Gamma_*^g$\\
	$L_x = \bigg\lbrace\left(\ket{g_0}\otimes\ket{+1}\right)\left(\bra{g_0}\otimes\bra{+1}\right) -$\\ \tab\tab\tab\tab  $\left(\ket{g_0}\otimes\ket{-1}\right)\left(\bra{g_0}\otimes\bra{-1}\right) \bigg\rbrace\oplus\mathbb{0}_2$
\end{flushleft}

\noindent
We then define the collapse operators along the intersystem crossing (ISC) pathway as follows:

\begin{flushleft}
	For the strong ISC occurring from the spin $\ket{\pm1}$ levels in $\ket{e_0}$ to $\ket{s_1}$,\\
	$\Gamma_x = \gamma_{es\pm1}$\\
	$L_x = \left\lbrace(\undertilde{0}_{n+3}\otimes\undertilde{0}_3)\oplus\ket{s_1}\right\rbrace  \left\lbrace(\bra{e_0}\otimes\bra{\pm1}) \oplus \undertilde{0}_2^\mathrm{T}\right\rbrace$
\end{flushleft}

\begin{flushleft}
	For the weak ISC occurring from the spin $\ket{0}$ level in $\ket{e_0}$ to $\ket{s_1}$,\\
	$\Gamma_x = \gamma_{es0}$\\
	$L_x = \left\lbrace(\undertilde{0}_{n+3}\otimes\undertilde{0}_3)\oplus\ket{s_1}\right\rbrace  \left\lbrace(\bra{e_0}\otimes\bra{0}) \oplus \undertilde{0}_2^\mathrm{T}\right\rbrace$
\end{flushleft}

\begin{flushleft}
	For the weak ISC occurring from $\ket{s_0}$ to the spin $\ket{\pm1}$ levels in $\ket{g_0}$,\\
	$\Gamma_x = \gamma_{sg\pm1}$\\
	$L_x = \left\lbrace(\ket{g_0}\otimes\ket{\pm1})\oplus\undertilde{0}_2\right\rbrace \left\lbrace (\undertilde{0}_{n+3}\otimes\undertilde{0}_3)^\mathrm{T} \oplus\bra{s_0}\right\rbrace$
\end{flushleft}

\begin{flushleft}
	For the strong ISC occurring from $\ket{s_0}$ to the spin $\ket{0}$ level in $\ket{g_0}$,\\
	$\Gamma_x = \gamma_{sg0}$\\
	$L_x = \left\lbrace(\ket{g_0}\otimes\ket{0})\oplus\undertilde{0}_2\right\rbrace \left\lbrace (\undertilde{0}_{n+3}\otimes\undertilde{0}_3)^\mathrm{T} \oplus\bra{s_0}\right\rbrace$
\end{flushleft}

\begin{flushleft}
	For the decay between the singlet states $\ket{s_1}$ and $\ket{s_0}$,\\
	$\Gamma_x = \gamma_s$\\
	$L_x = (\mathbb{0}_{n+3}\otimes\mathbb{0}_3)\oplus\ket{s_0}\bra{s_1}$
\end{flushleft}
\vspace{-1em}

\subsection{Plasmonic modification of optical decay rates and absorption}\label{Sec:Plasmon_decay_rates_&_absorption}
We use the equations derived by Carminati \emph{et al.} \cite{carminati2006radiative} (extended for a generic background medium) to estimate the decay rate modifications experienced by an NV centre within the classical dipole approximation. 

We consider cases where the NV optical dipoles are oriented normal ($\perp$) and tangential ($\parallel$) to the surface of an adjacent spherical plasmonic nanoparticle. In our context, $(\gamma_{k,m})_\perp$ and $(\gamma_{k,m})_\parallel$ below represent the respective plasmon-modified total emission rates of the $k^\text{th}$ NV optical transition in a medium of refractive index $n_\text{b}$. These rates accompany the emission Lindblad operators discussed in the earlier section. 

The absorption of the plasmonic nanoparticle could be treated as a nonradiative contribution, with rates $(\gamma_{k,m}^\text{\tiny{NR}})_\perp$ and $(\gamma_{k,m}^\text{\tiny{NR}})_\parallel$, in the context of far field emission. These rates, the contributions of which are already included in the total emission rates, are used when estimating the quantum efficiencies.

\begin{widetext}
	\begin{subequations}\label{Eq:Decay_near_large_MNPs}
		\begin{align*}
			\frac{(\gamma_{k,m})_\perp}{\gamma_{k,m}^\text{f}} &\approx n_\text{b}\left\lbrace 1 + 6 k_\text{b}^3 \mathrm{Im}\left[\alpha(\omega) e^{2ik_\text{b}R}\left(\frac{-1}{(k_\text{b}R)^4} + \frac{2}{i(k_\text{b}R)^5} + \frac{1}{(k_\text{b}R)^6}\right)\right]\right\rbrace, \\
			\frac{(\gamma_{k,m})_\parallel}{\gamma_{k,m}^\text{f}} &\approx n_\text{b}\left\lbrace 1 + \frac{3}{2}k_\text{b}^3\mathrm{Im}\left[\alpha(\omega)e^{2ik_\text{b}R}\left(\frac{1}{(k_\text{b}R)^2} - \frac{2}{i(k_\text{b}R)^3} - \frac{3}{(k_\text{b}R)^4} + \frac{2}{i(k_\text{b}R)^5} +\frac{1}{(k_\text{b}R)^6}\right)\right] \right\rbrace,\\\\
			\frac{(\gamma_{k,m}^\text{\tiny{NR}})_\perp}{\gamma_{k,m}^\text{f}} &\approx 6 n_\text{b} k_\text{b}^3 \left\lbrace\mathrm{Im}\left[\alpha(\omega)\right] - \frac{2}{3}k_\text{b}^3 |\alpha(\omega)|^2\right\rbrace \left\lbrace \frac{1}{(k_\text{b}R)^6} + \frac{1}{(k_\text{b}R)^4} \right\rbrace, \\
			\frac{(\gamma_{k,m}^\text{\tiny{NR}})_\parallel}{\gamma_{k,m}^\text{f}} &\approx \frac{3}{2}n_\text{b}k_\text{b}^3\left\lbrace\mathrm{Im}\left[\alpha(\omega)\right] - \frac{2}{3}k_\text{b}^3 |\alpha(\omega)|^2\right\rbrace \left\lbrace \frac{1}{(k_\text{b}R)^6} - \frac{1}{(k_\text{b}R)^4} + \frac{1}{(k_\text{b}R)^2} \right\rbrace,
		\end{align*}
	\end{subequations}
\end{widetext}
In the above equations, the free-space optical decay rate of the respective emission transition is denoted by $\gamma_{k,m}^\text{f}$, and $\omega$ is the angular frequency of emission. 

We define the fraction of emitted photons surviving in the far field following plasmonic absorption as the relative quantum efficiency ($Q_{k,m}$) of a given transition, assuming near-unity quantum efficiency for the isolated NV centre. This is estimated for each emission band as,
\begin{equation*}
	Q_{k,m} = \left(\gamma_{k,m} - \gamma_{k,m}^\text{\tiny{NR}}\right)\big/\gamma_{k,m}.
\end{equation*}

\subsection{Emission intensity spectra}
Utilising the steady state density matrix obtained by solving the master equation outlined earlier, we estimate the far-field emission intensity spectrum of the NV centre as follows \cite{hapuarachchi2022nv, campaioli2023tutorial},
\begin{subequations}
\begin{align*}\label{Eq:Total_emission_intensity}
	I_\text{\tiny{NV}}(\omega) &\propto \sum_k\sum_m Q_{k,m}\gamma_{k,m} \int_{-\infty}^{\infty}d\tau e^{-i\omega\tau}C_{k,m}(\tau, 0),\\
	&\text{where,} \;\;C_{k,m}(\tau, 0) = \langle\sigma_{k,m}^\dagger(\tau)\sigma_{k,m}(0)\rangle_\text{ss}.
\end{align*}
\end{subequations}
The operator $\langle\cdot\rangle_\text{ss}$ denotes expectation calculated using the steady state density matrix. 

When calculating the near-field NV emission spectra in the presence of a plasmonic nanoparticle and near/far field emission spectra of isolated NV centres, we set $Q_{k,m}=1$ $\forall \; k,m$. The earlier expression was used to compute the emission intensity spectra compared with existing experiments, in the section \ref{Sec:Supp_results}.

\section{Further information on simulations}

\subsection{Common parameters}

In this work, we use the set of NV parameters reported by Albrecht \emph{et al.} in \cite{albrecht2013coupling} for ground state phonon (vibronic) energies $\hbar\omega_{k,m}$, free-space decay rates $\gamma_{k,m}^\text{f}$, and ground state vibronic decay rates $\gamma_{k,k-1}$ of a single NV centre residing in a nanodiamond at room temperature. These parameters are presented in table\ \ref{Table:Params_from_Roland}. We modify the free-space NV decay rates $\gamma_{k,m}^\text{f}$ to obtain the plasmonically modified rates $\gamma_{k,m}$ as outlined in section \ref{Sec:Plasmon_decay_rates_&_absorption}. We assume that the parameters in table \ref{Table:Params_from_Roland} apply for all three spin states of a given $k$, and that the respective transitions are spin preserving. 

\begin{table}[h!]
	\begin{center}
		\begin{tabular}{ |c|c|c|c|c| } 
			\hline
			k & \makecell{$\hbar\omega_{k,m}$ \\ $(\SI{}{\milli\electronvolt})$} & \makecell{$\gamma_{k,m}^\text{\tiny{f}}$\\$(\SI{}{\mega\hertz})$} & \makecell{$\gamma_{k,k-1}$\\$(\SI{}{\tera\hertz}$)} \\ 
			\hline
			0 & 0     & 0.69  & -  \\ 
			1 & 31.8  & 2.42  & 85 \\ 
			2 & 70.3  & 8.57  & 82 \\
			3 & 124   & 7.57  & 79 \\ 
			4 & 168   & 6.46  & 88 \\ 
			5 & 221   & 4.23  & 65 \\
			6 & 275   & 3.03  & 71 \\ 
			7 & 319   & 1.51  & 86 \\
			\hline
		\end{tabular}
		\caption{Ground state vibronic energies $\hbar\omega_{k,m}$, free-space decay rates $\gamma_{k,m}^\text{f}$, and ground state vibronic decay rates $\gamma_{k,k-1}$ of a single NV centre in a nanodiamond at room temperature from \cite{albrecht2013coupling}. Here we assume that the same parameters apply for all spin states of a given $k$.}
		\label{Table:Params_from_Roland}
	\end{center}
\end{table}

Throughout this work, we use energy of the NV zero-phonon line $\hbar\omega_\text{z} = \SI{1.941}{\electronvolt}$ \cite{albrecht2013coupling}, energy of the phenomenological upper excited level $\ket{e_1}$ resonant with the incoming optical field $\hbar\omega_\text{d}\approx\SI{2.033}{\electronvolt}$ (which converts to a free-space optical wavelength of $\sim$$\SI{532}{\nano\meter}$), an optical dephasing rate $\gamma_*=\SI{15}{\tera\hertz}$ \cite{albrecht2013coupling}, and $n_\text{D}\sim2.4$. Using the information of the confocal setup and the effective incoherent pumping rate reported in \cite{albrecht2013coupling}, we estimate an excited state vibronic decay rate $\gamma_\text{e}\sim\SI{1434}{\tera\hertz}$ for an NV centre in a nanodiamond at room temperature. 

We approximate the dipole moment elements $\mu_{e_{jm}g_{0m}}$ that correspond to the $|e_j\rangle\leftrightarrow|g_0\rangle$ transitions (for both $j=0$ and $1$, $\forall m$) with the absolute angle-averaged optical dipole moment element value $\SI{5.2}{D}$ reported in \cite{alkauskas2014first}. We estimate the dipole moment elements for the other transitions $|e_j\rangle\leftrightarrow|g_k\rangle$ ($\forall m$) as $\mu_k=\mu_0\sqrt{\gamma^f_{k,m}/\gamma^f_{0,m}}$, such that $\gamma_k^\text{f} \propto |\mu_k|^2$ for each transition \cite{fox2006quantum, carmichael1999statistical, premaratne2021theoretical, hapuarachchi2022nv}. 

We assume that the plasmon excitation occurs at the angular frequency of the incoming optical radiation $\omega_\text{d}$, and the dielectric permittivity of the plasmonic particles $\epsilon_\text{m}(\omega_\text{d})$ are obtained by interpolating the tabulations by Johnson and Christy \cite{johnson1972optical}. 

The parameters related to the spin subsystems and the intersystem crossing pathway of the NV centre are presented in table \ref{Table:Spin_and_ISC} below. The values of any other parameters specific to a given figure can be found in the respective caption.
{\renewcommand{\arraystretch}{1.3}
\begin{table}[h!]
	\begin{center}
		\begin{tabular}{ |c|c|c| } 
			\hline
			Parameter & Description & \makecell{Value and \\source} \\ 
			\hline
			$\gamma_{es\pm1}$     	& \makecell{ISC from $\ket{e_0}\otimes\ket{\pm1}\to\ket{s_1}$}  	& $\SI{92}{\mega\hertz}$ \cite{gupta2016efficient}  \\ 
			$\gamma_{es0}$  		& \makecell{ISC from $\ket{e_0}\otimes\ket{0}\to\ket{s_1}$}  		& $\SI{11.4}{\mega\hertz}$ \cite{gupta2016efficient} \\ 
			$\gamma_{sg\pm1}$  		& \makecell{ISC from $\ket{s_0}\to\ket{g_0}\otimes\ket{\pm1}$}  	& $\SI{2.35}{\mega\hertz}$ \cite{gupta2016efficient} \\
			$\gamma_{sg0}$   		& \makecell{ISC from $\ket{s_0}\to\ket{g_0}\otimes\ket{0}$}  		& $\SI{4.84}{\mega\hertz}$ \cite{gupta2016efficient} \\ 
			$\gamma_{s}$   			& \makecell{Total decay from $\ket{s_1}\to\ket{s_0}$}  				& $\SI{1}{\giga\hertz}$ \cite{acosta2010optical}\\ 
			$\Delta E_s$   			& \makecell{Energy between $\ket{s_1}$ and $\ket{s_0}$}  			& $\SI{1.19}{\electronvolt}$ \cite{acosta2010optical, kim2021absorption, wolf2023nitrogen}\\
			$\Delta E_{es}$   		& \makecell{Energy between $\ket{e_0}$ and $\ket{s_1}$}  			& $\SI{0.4}{\electronvolt}$ \cite{attrash2023high}\\ 
			$\Gamma_\text{rel}^g$   & \makecell{Ground state spin relaxation} 							& $\frac{1}{\SI{7.7}{\milli\second}}$ \cite{Susumu2008Quenching, doherty2013nitrogen}\\
			$\Gamma_\text{rel}^e$   & \makecell{Excited state spin relaxation}  						& $\frac{1}{\SI{1}{\milli\second}}$ \cite{busaite2020dynamic}\\
			$\Gamma_*^g$   			& \makecell{Ground state spin dephasing}  							& $\frac{1}{\SI{6.7}{\micro\second}}$ \cite{Susumu2008Quenching, doherty2013nitrogen}\\
			$\Gamma_*^e$   			& \makecell{Excited state spin dephasing}  							& $\frac{1}{\SI{10}{\nano\second}}$ \cite{busaite2020dynamic}\\
			\hline
		\end{tabular}
		\caption{Parameters related to the NV spin subsystems and the intersystem crossing (ISC) pathway used throughout this work, and the sources.}
		\label{Table:Spin_and_ISC}
	\end{center}
\end{table}
\vspace{-1em}

\subsection{Supplementary results}\label{Sec:Supp_results}

Further verifications of our theoretical formalism against existing experiments in the literature can be found in Fig.\ \ref{Fig:SuppMat_Verification}. Here we compare the output of our model against the fully optical emission spectra measured from isolated NV centre in a nanodiamond in air, at room temperature, by Albrecht \emph{et al.} in \cite{albrecht2013coupling}. We then perform similar comparisons in the presence of a plasmonic nanoparticle using the measurements reported by Schietinger \emph{et al.} based on their controlled NV-plasmonic coupling study in \cite{schietinger2009plasmon}. 

From the above comparisons and those presented in the main text, it is evident that our model is capable of successfully capturing the experimentally expected operation of an NV centre, both in the presence and absence of plasmonic interactions.

Finally, in Fig.\ \ref{Fig:MNP_based_plots_suppmat}, we present the plasmonically modified properties of the NV centre, considered when selecting the parameter region explored in the main text. These include the spectral variations of plasmonically modified optical Rabi frequency, decay rate modification, and relative quantum efficiency, for an NV centre considered as a generic quantum emitter.

\begin{figure*}[t!]
	\includegraphics[width=\textwidth]{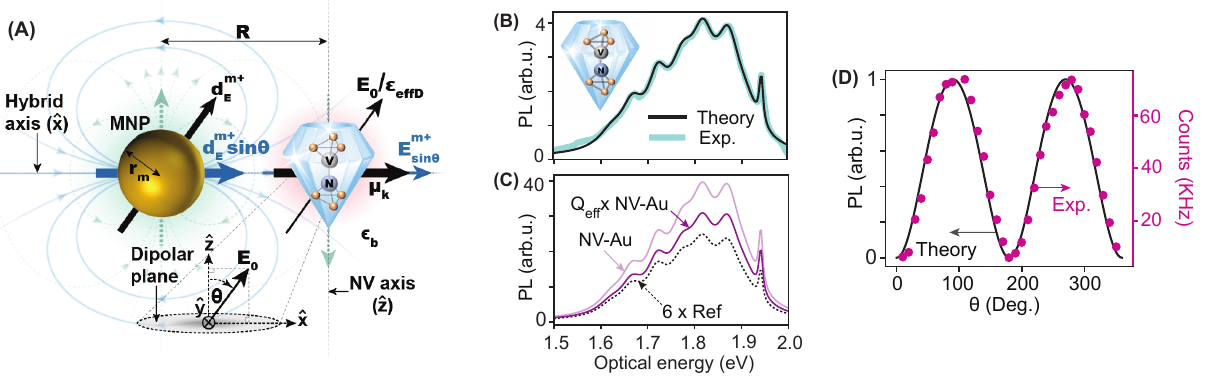}
	\centering
	\caption{Verification of the open quantum system model against fully optical experiments in the literature. (A) Abstract illustration of the experimental NV-Au dimer setup of Schietinger \emph{et al.} in \cite{schietinger2009plasmon}. Here, the optical excitation is polarized along a plane perpendicular to the NV dipolar plane that passes through the NV axis and the hybrid axis. $\bm{E}_0$ and $\bm{d}_\text{\tiny{E}}^\text{m+}$ represent the positive frequency amplitudes (coefficients of $e^{-i\omega_\text{\tiny{d}}t}$) of the external optical field and the resulting AuNP dipole field, respectively. $E_0 \sin\theta/\epsilon_\text{effD}$ and $\bm{E}^\text{m+}_{\sin\theta}$ represent the screened projections of $\bm{E}_0$ and AuNP dipole response field onto the NV dipolar plane, at the NV location. The AuNP radius $r_\text{m}\sim\SI{30}{\nano\meter}$, estimated centre separation $R\sim\SI{38}{\nano\meter}$ \cite{hapuarachchi2022nv}, background permittivity $\epsilon_\text{b}\sim 1$ (air), and we assume an optical excitation intensity $I\sim\SI{0.5}{\milli\watt\per\micro\meter\squared}$. (B) Verification of the simulated optical emission of an isolated NV centre in a nanodiamond against the experimental measurements by Albrecht \emph{et al.} in \cite{albrecht2013coupling}, assuming an optical excitation intensity $I\sim\SI{0.5}{\milli\watt\per\micro\meter\squared}$. (C) Spectra at the top, middle, and bottom depict the near-field NV emission spectrum of the NV centre in (A) when $\theta=\pi/2$ radians, the respective far-field spectrum (obtained by scaling the near-field spectrum by the effective quantum efficiency $\sim$0.78 reported in \cite{schietinger2009plasmon}), and the spectrum of the isolated NV centre magnified six times, respectively. It is evident that the model is capable of successfully reproducing the $\sim$6-fold emission enhancement near the NV ZPL experimentally reported in \cite{schietinger2009plasmon}. (D) Comparison of the variation of total PL against $\theta$ in the setup shown in (A) against the experimental measurements reported by Schietinger \emph{et al.} \cite{schietinger2009plasmon}.}\label{Fig:SuppMat_Verification}
\end{figure*}

\begin{figure*}[t!]
	\includegraphics[width=\textwidth]{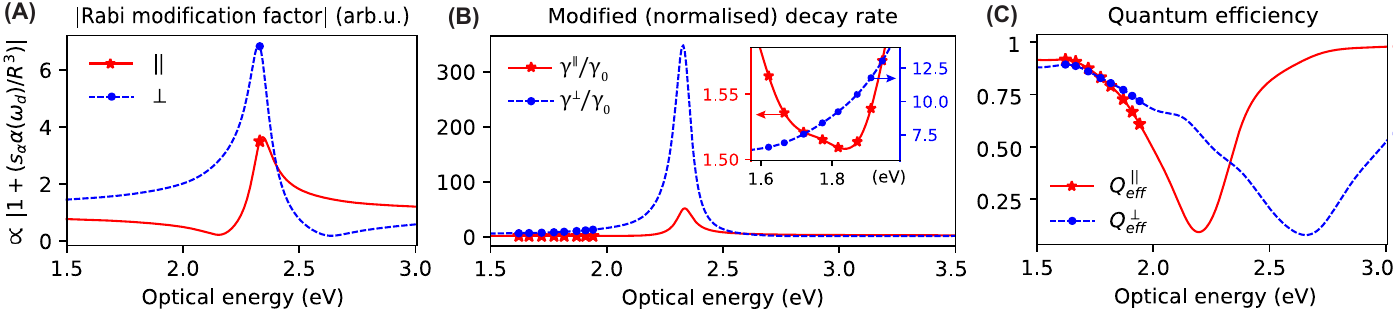}
	\centering
	\caption{Plasmonically modified properties of the NV centre as an emitter, considered when selecting the parameter region explored in the main text. (A) Line shape of the absolute value of plasmonically induced optical Rabi frequency modification for the NV$^\parallel$Ag and NV$^\perp$Ag setups (where the AgNP radius $r_\text{m}=\SI{10}{\nano\meter}$ and centre separation $R=\SI{20}{\nano\meter}$) presented in Fig 3 of main text. The blue dot and red star depict the respective values at the AgNP's plasmonic resonance, which is selected as the optical illumination frequency for both setups. (B) Decay rate enhancements experienced by a classically modelled dipolar emitter, at different emission frequencies, due to the above AgNP in the configurations where the emitter dipole is tangential ($\parallel$) and radial ($\perp$) to the AgNP surface. Normalisation considers the free space decay rate $\gamma_0$ of the same emitter. The blue dots and red stars depict the rate enhancements at the NV emission band peaks, for the two configurations. (C) Quantum efficiency ($Q_\text{eff}$) defined as the proportion of emitted photons surviving in the far field following absorption by the adjacent AgNP, for emitters in $\parallel$ and $\perp$ configurations. $Q_\text{eff}$ values at the NV emission band peaks are depicted in blue circles and red stars.} \label{Fig:MNP_based_plots_suppmat}
\end{figure*}

\clearpage

\bibliographystyle{unsrt}

\providecommand{\noopsort}[1]{}\providecommand{\singleletter}[1]{#1}%


\end{document}